\documentclass[
aps,%
11pt,%
final,%
notitlepage,%
oneside,%
onecolumn,%
nobibnotes,%
nofootinbib,%
superscriptaddress,%
noshowpacs,%
centertags]%
{revtex4}

\begin{document}
\title{MISSING DARK MATTER IN THE LOCAL UNIVERSE}

\author{\firstname{I.~D.}~\surname{Karachentsev}}
\affiliation{\saoname}


\begin{abstract}
A sample of 11 thousand galaxies with radial velocities $V_{\rm
LG}<3500$~km/s is used to study the features of the local
distribution of luminous (stellar) and dark matter within a sphere
of  radius of around 50~Mpc around us. The average density of
matter in this volume, $\Omega_{m,{\rm loc}}=0.08\pm0.02$, turns
out to be much lower than the global cosmic density
$\Omega_{m,{\rm glob}}=0.28\pm0.03$. We discuss three possible
explanations of this paradox: 1) galaxy groups and clusters are
surrounded by extended dark halos, the major part of the mass of
which is located outside their virial radii; 2) the considered
local volume of the Universe is not representative, being situated
inside a giant void; and 3) the bulk of matter in the Universe is
not related to clusters and groups, but is rather distributed
between them in the form of massive dark clumps. Some arguments in
favor of the latter assumption are presented. Besides the two
well-known inconsistencies of modern cosmological models with the
observational data: the problem of missing satellites of normal
galaxies and the problem of missing baryons, there arises another
one---the issue of missing dark matter.


\end{abstract}
\maketitle


\section{INTRODUCTION}

Different cosmological models are commonly tested based on the
observations of distant objects with redshifts of $z=v/c\sim1$ and
properties of the cosmic background radiation generated at the
epoch of $z\sim10\,000$. However, the structure and kinematics of
nearby $(z\simeq0)$ volumes of the Universe   is also an important
source of cosmological data.  In the Local universe,
conditionally limited within the radius of $D=10$~Mpc,   a large
number of dwarf galaxies was found, the velocities and distances
of which are tracing the Hubble flow with an unprecedented detail,
as compared with distant objects. The studies of stellar
population of nearby galaxies allow to recover the history of star
formation in them with a resolution of $\Delta
T\sim10^8$--$10^9$~yr. In fact, over the past \mbox{10--15}~years
the study of the Local universe has become an independent and
fruitful branch of observational cosmology, which has been
repeatedly emphasized by
Peebles~\cite{Peebles1993:Karachentsev,Peebles2001:Karachentsev,Peebles2010:Karachentsev}.

\begin{figure*}[tbp!!!]
\setcaptionmargin{5mm}
\onelinecaptionstrue
\includegraphics[width=\textwidth]{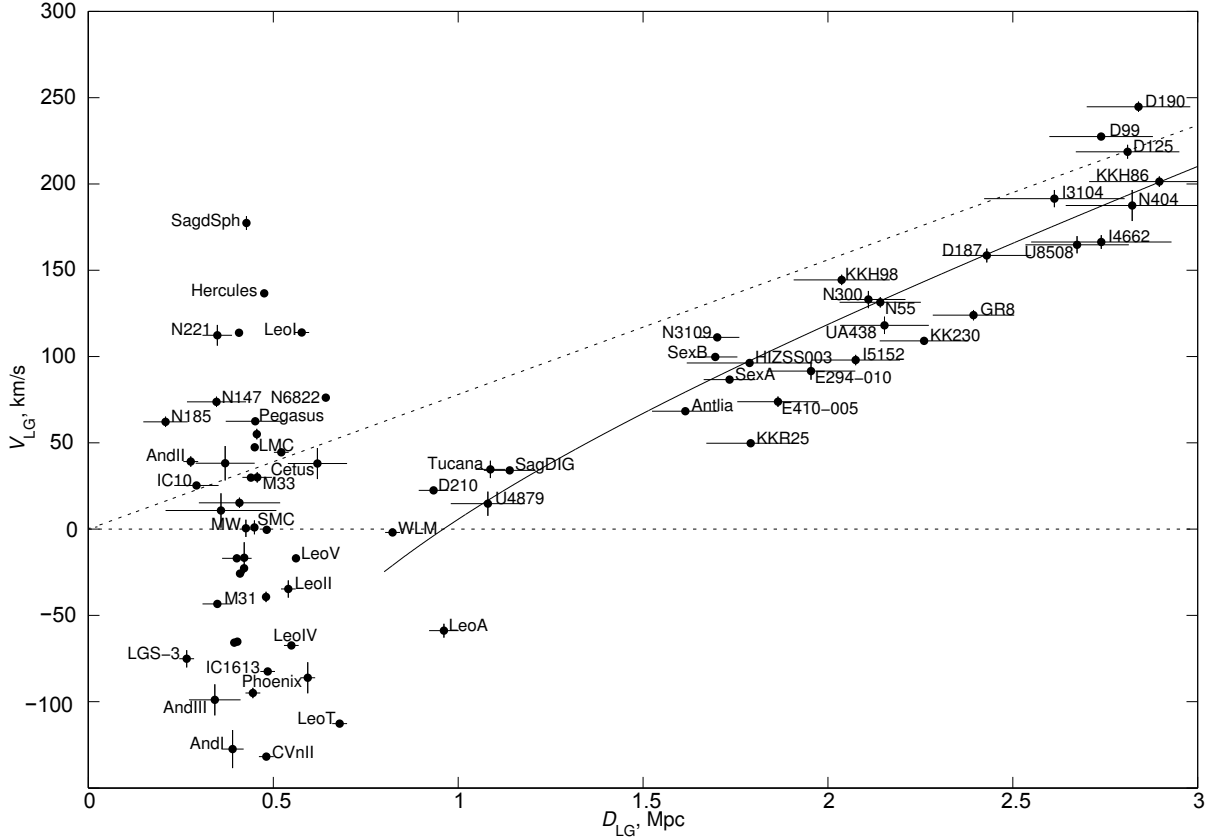}
\captionstyle{normal} \caption{The Hubble diagram for galaxies
in the vicinity of the Local Group. The distances and velocities
of galaxies are given with respect to the centroid of the Local
Group, supplemented with error bars. The dashed line corresponds
to the undistorted Hubble flow with the parameter
$H_0=80$~km/s/Mpc. The curved solid line describes the dragging
effect on the flow by the Local Group with a mass of
$1.9\times10^{12}~M_{\odot}$.}
\end{figure*}

Until recently, the scarcity of data on the distances of even the
closest galaxies was the major obstacle in the development of
observational cosmology in the Local universe. Deployment of
unique capabilities of the Hubble Space Telescope, combined with a
new method for determining the distances to galaxies by the
luminosity of the tip of the red giant branch
(TRGB)~\cite{Lee1993:Karachentsev} allowed to carry out mass
distance measurements of more than 250 nearby galaxies with an
accuracy of 5--10\%. The summary of data on distances, radial
velocities and other parameters of galaxies in the Local Volume
within a radius of 10~Mpc was presented in the Catalog of
Neighboring Galaxies (CNG, \cite{Kar2004:Karachentsev}), which
contains 450 objects. In this volume, where the dwarf galaxies
down to the luminosities of 10\,000~times lower than that of the
Milky Way are visible, there are more than a dozen groups, similar
to our Local Group  in size and population. A detailed pattern of
motions of galaxies in these groups and around them has for the
first time revealed unexpected features in the Hubble flow at
small \mbox{(1--3~Mpc)} scales. It turned out that the Hubble
velocity--distance diagrams around the Local Group and other
neighboring groups are characterized by a small dispersion of
peculiar velocities $V_{\rm pec} \sim30$~km/s. With such small
chaotic motions and minor distance measurement errors, a
distortion of the ``cold'' Hubble flow becomes noticeable, caused
by the gravitational drag of the galaxies, surrounding the group,
by the total mass of the group itself.

An example of the Hubble flow around our group is demonstrated in
 Fig.~1. This diagram exhibits the region of virial motions of $\pm$200~km/s
in the companions of the Milky Way  and M~31 (Andromeda) and the
region of the total Hubble expansion.  They are separated from
each other by the ``zero-velocity sphere'' with the radius of
\mbox{$R_0=(0.96\pm0.03)$~Mpc \cite{Kar2009:Karachentsev}}. It is
noteworthy that the radius $R_0$ determines the total mass of the
group, \mbox{$M_T=(1.9\pm0.2)\times10^{12}~M_{\odot}$}, \linebreak
and this value is in a remarkable agreement \linebreak with the
virial mass estimates of \linebreak \mbox{$M({\rm MW}+{\rm M\,31})
= (1.6\div2.2)\times10^{12}~M_{\odot}$}. It should be emphasized
that the listed value of $M_T$ is obtained in assumption of the
standard  $\Lambda$CDM cosmological model with the parameter
$\Omega_{\lambda}=0.73$. In the absence of the \mbox{$\lambda$
term} ($\Omega_{\lambda}=0$), the estimate of the total mass via
external galaxy motions would only amount to
\mbox{$1.2\times10^{12}~M_{\odot}$,} i.e. lower than the   virial
estimates. A similar conclusion can be drawn from the motions of
galaxies around the neighboring groups, which are dominated by the
M\,81 and Centaurus\,A galaxies~\cite{Kar2005:Karachentsev}.
Therefore, the observed features of the local Hubble flow give (at
the recently achieved distance measurement accuracy) a direct and
independent evidence of the presence in the Universe of a specific
cosmic component, the dark energy, discovered from the
observations of distant Supernovae.

Recent massive sky surveys both in the optical range, and at the
radio line wavelength of 21~cm led to the discovery of numerous
nearby dwarf galaxies. An updated version of the CNG catalog now
consists of  about 800 objects and is currently prepared for
\linebreak print \cite{Kar2012:Karachentsev}.

\section{FROM  LOCAL VOLUME, $D<10$ MPC,  TO LOCAL UNIVERSE, $D<50$ MPC}

\begin{figure*}[tbp!!!]
\setcaptionmargin{5mm} \onelinecaptionstrue
\includegraphics[scale=0.8]{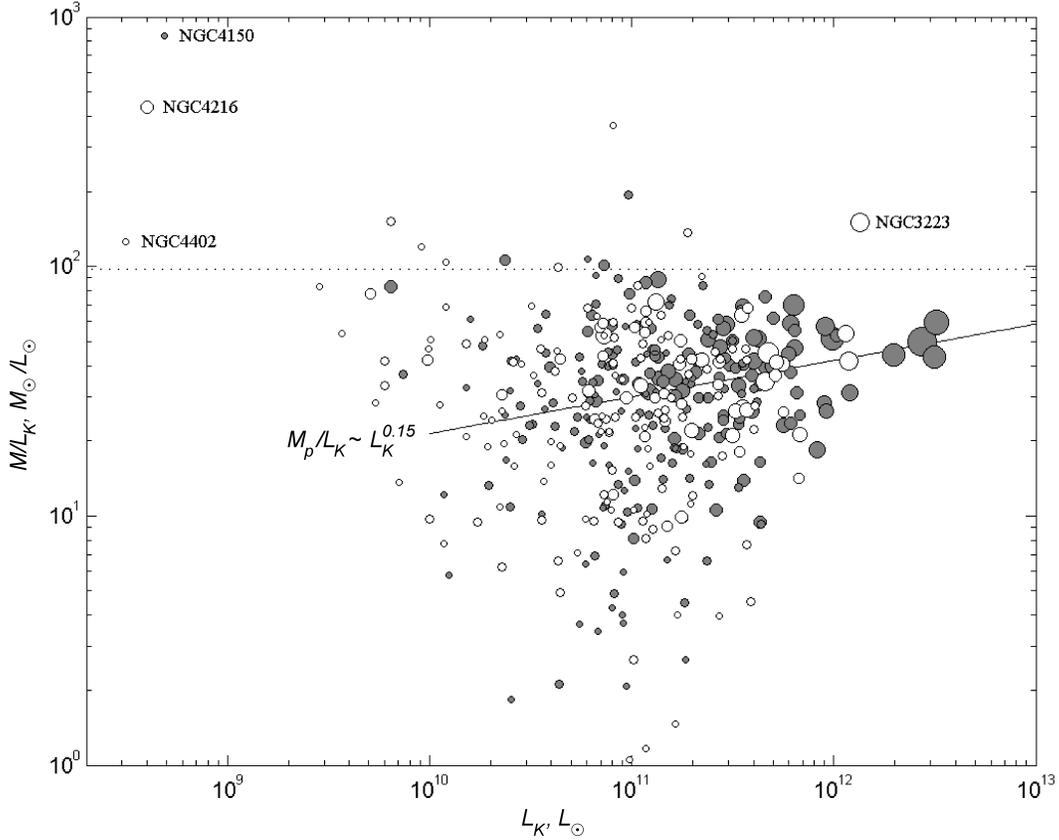}
\captionstyle{normal} \caption{The distribution of groups of
galaxies with  4 or more members by the integral $K$-band
luminosity and the virial mass-to-luminosity ratio. The size of
the circle depicts the population of the system. Dark circles mark
the groups and clusters, where the brightest galaxy has an early
morphological type (E, S0, Sa).}
\end{figure*}

High density of observational data on the galaxies in the Local
Volume  allows us to have a fairly complete conception of the
spatial distribution of  luminous and dark matter in it. However,
on the scale of  $D\sim10$~Mpc  relative fluctuations of the
luminosity density amount to
$\Delta\rho_L/\overline{\rho_L}\sim1$. This is why the Local
Volume cannot be considered representative in terms of its
kinematics and dynamics. To achieve a better representation of the
observational sample, Makarov and Karachentsev
\cite{Mak2011:Karachentsev} have examined a 100~times larger
volume of space around our Galaxy.  This volume includes all known
galaxies with radial velocities $V_{\rm LG}<3500$~km/s relative to
the centroid of the Local Group after deduction of the zone of
strong extinction at Galactic latitudes  \mbox{$\mid
b\mid<15^{\circ}$}.  This volume with a diameter of $96$~Mpc (at
\mbox{$H_0=73$~km/s/Mpc}) contains about 11 thousand galaxies.
Most of them belong to the Local Supercluster, but this volume
(called the ``Local universe'') embraces the ridges of other
neighboring superclusters. Based on the data of the Sloan Digital
Sky Survey \cite{Aba2009:Karachentsev}, Papai and \mbox{Szapudi
\cite{Pap2010:Karachentsev}} have estimated that the variations of
luminosity in galaxies in the cube with an edge of $100$~Mpc are
about 10\%. Consequently, the considered volume of Local universe
satisfies quite well the condition of sample representativeness.

The main efforts in our program were aimed at the systematization
of data on radial velocities, apparent magnitudes and
morphological types of galaxies. At that, we made a search for new
dwarf galaxies and performed optical identification of H\,I radio
sources from the HIPASS  \cite{Meyer2004:Karachentsev}, ALFALFA
\cite{Gio2005:Karachentsev}   and other ``blind'' sky surveys.
Particular attention was paid to the problem of the so-called
``astro-spam'': frequent cases of confusion with the
identification of H\,I sources, the superposition of a star over
a galaxy image, when a nearly zero velocity of the star
was attributed to a distant galaxy, the cases of a false
multiplicity of  galaxies, when two or more clumps in it
were mistaken for   dynamically different objects, etc.

\begin{figure*}[tbp!!!]
\setcaptionmargin{5mm} \onelinecaptionstrue
\includegraphics[scale=0.8]{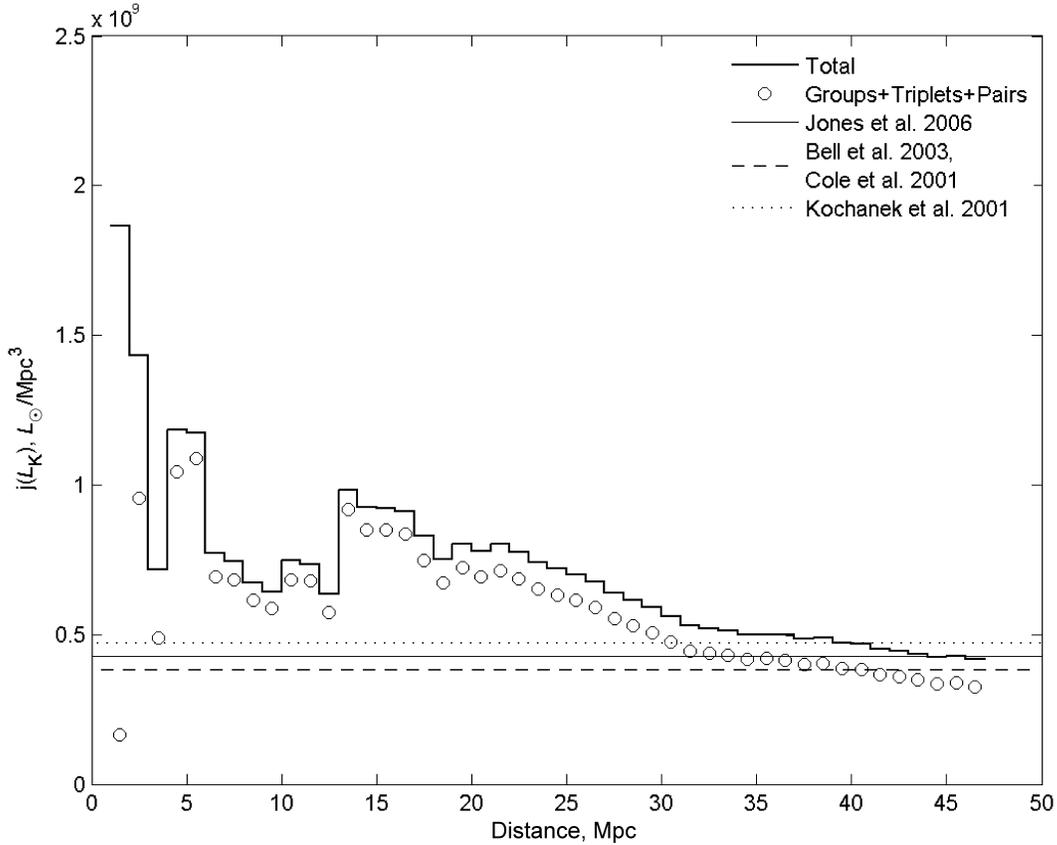}
\captionstyle{normal} \caption{The mean density of $K$-luminosity
of the Local universe in the spheres of different radii (the
stepped line).  The circles mark the contribution of galaxies in
groups, triplets and pairs. The four horizontal lines fix the
global value of the luminosity density according to different
authors.}
\end{figure*}

Makarov and Karachentsev  \cite{Mak2011:Karachentsev} have applied
a new algorithm for identifying the groups with different
populations $n$ to the updated and ``cleaned'' sample of 10\,900~
galaxies. In contrast to the simple ``friends of friends'' (FoF)
percolation algorithm \cite{Huch1982:Karachentsev}, the grouping
criterion we used  took into account the individual luminosities
of galaxies. At the primary stage of combining two given galaxies
in a virtual pair it was assumed that this pair should have a
negative total energy and its components must be causally related
(the ``crossing  time'' of a pair is shorter than the age of the
Universe). As shown by the subsequent analysis, used in
\cite{Mak2011:Karachentsev}, the algorithm isolates groups with
approximately the same characteristics both in the nearby, and
distant volumes of the Local universe. (Note that the above
mentioned  FoF criterion does not possess this property). As a
result, the following catalogs were compiled: 509~pairs
\cite{Kar2008:Karachentsev}, 168~triple systems
\cite{Mak2009:Karachentsev}, 395~groups with the population of
\mbox{$n>3$} \cite{Mak2011:Karachentsev} and 520~very isolated
\mbox{galaxies \cite{Kar2011b:Karachentsev}}.

Not going into detail,  note that the pairs, groups and clusters
in the Local universe, selected by this algorithm are in good
correspondence with the previously known systems. In some cases,
however, the new criterion breaks the known galaxy aggregates into
subsystems, which are obviously associated with each other, but
have not yet reached the stage of merging and relaxation. Ignoring
the substructure of such entities markedly overstates the estimate
of their virial mass.

The distribution of groups of galaxies with a population of
\mbox{$n>3$} by the integral \mbox{$K$-band}
luminosity and the virial (projected) mass-to-luminosity ratio is
shown in Fig.~2~\cite{Mak2011:Karachentsev}. The groups, where the
main galaxy belongs to the early \mbox{($T<3$)} or late types, are
represented by shaded and empty circles, respectively, the sizes
of the circles are proportional to the group population.
The dashed horizontal line in the figure at
\mbox{$M/L_K=97~M_{\odot}/L_{\odot}$} corresponds to the global
average  density of matter \mbox{$\Omega_m=0.28$.}

A considerable scatter of galaxy systems in this diagram is
primarily due to the projection factors. Despite the large
variations, the average   virial mass-to-luminosity ratio grows
with an increase in the system population or luminosity and is
correlated with the morphological type of the brightest member.
All these features are well known from other catalogs of groups
and clusters of galaxies.

\begin{figure*}[tbp!!!]
\setcaptionmargin{5mm} \onelinecaptionstrue
\includegraphics[scale=0.8]{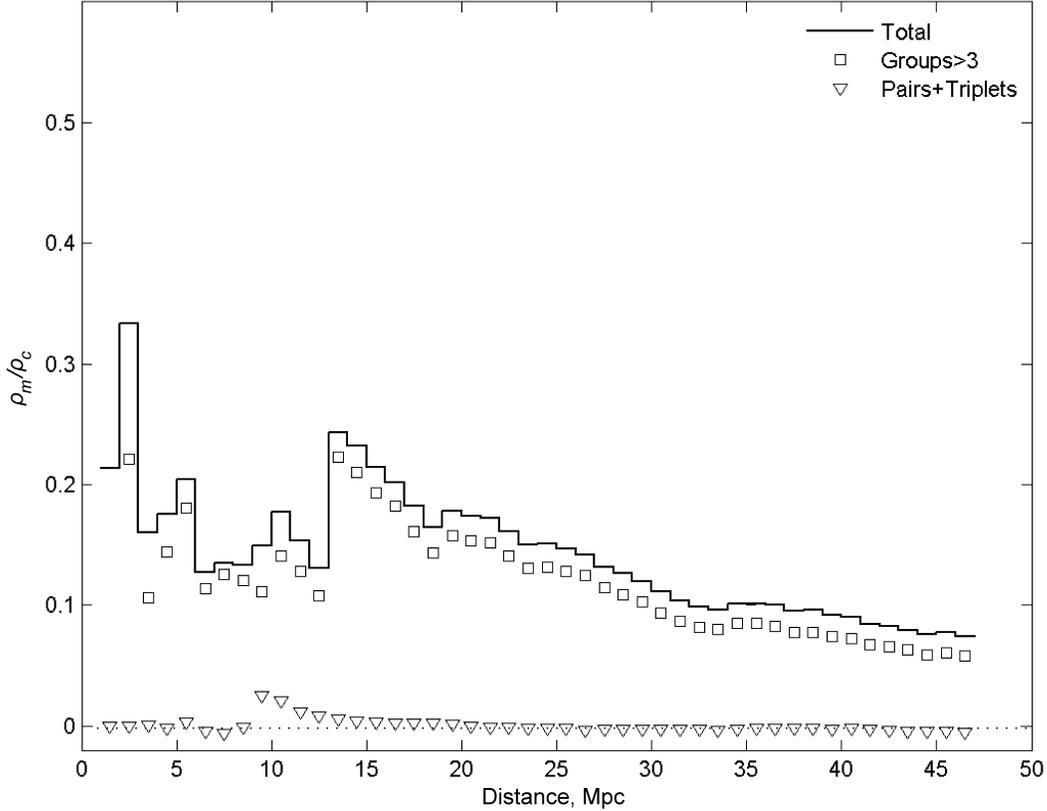}
\captionstyle{normal} \caption{The average density of matter in
the spheres of different radii (the stepped line). The squares and
triangles mark the contribution of pairs, triplets, and groups of
galaxies.}\end{figure*}

Possessing the data on the luminosities of galaxies in the Local
universe, Makarov and Karachentsev~\cite{Mak2011:Karachentsev}
have traced the  variation of the mean luminosity  density  in the
$K$-band depending on the radius of the sphere, within which the
averaging is carried out. The dependence of the mean density
\mbox{$j(L_K)$} on distance $D$ is presented in Fig.~3 with a step
of 1~Mpc.  As we can see, the main contribution to the total
luminosity at all scales is given by the   members of groups,
triple systems and pairs (circles). Taking into account the
contribution of  non-clustered galaxies, the total luminosity
density (the stepped line) gradually approaches the asymptotic
value, four estimates of which according to  the 2MASS
\mbox{survey
\cite{Cole2001:Karachentsev,Bell2003:Karachentsev,Kochan2001:Karachentsev,Jon2006:Karachentsev}}
are marked with horizontal lines.  At all scales, the local
luminosity density exceeds the global value. A broad hump at the
distances of \mbox{15--25~Mpc} is due to the contribution of the
Virgo and Fornax clusters. Since in the $K$-band the mean stellar
mass-to-luminosity ratio of galaxies is
\mbox{$M_*/L_K\simeq1~M_{\odot}/L_{\odot}$
\cite{Bell2003:Karachentsev}}, Fig.~3   shows an excess of stellar
mass density over its global value at all scales of the Local
universe up to the boundary of the sample
\mbox{($D\simeq45$~Mpc)}.

Summing up the virial masses of groups and clusters, Makarov and
Karachentsev \cite{Mak2011:Karachentsev} have built the
distribution of the mean density of dark matter in the spheres of
various radii around our Galaxy. The results are demonstrated in
Fig.~4.  The contribution of groups with the population of $n>3$
is shown by squares, the contribution of pairs and triplets---by
triangles, and the stepped line represents the course of the total
mean density of matter, taking into account the non-clustered
``field'' galaxies, the relative contribution of which by
population is 46\%, and by luminosity---18\% (the field galaxies
are dominated by the dwarf population).  The mass-to-luminosity
ratio of the non-clustered galaxies was taken here to be the same
as for pairs and triple systems, i.e. about
$20~M_{\odot}/L_{\odot}$,  which is consistent with the estimates
of    $M/L_K$ from the effect of weak gravitational lensing
(see~\cite{Man2006:Karachentsev,van2011:Karachentsev}).

As follows from the comparison of Figs.~4 and 3, the course of the
mean   density of dark matter with   distance $D$ approximately
repeats the course of the mean stellar mass density. However, in
almost all bins the density \mbox{$\Omega_m=\rho_m/\rho_c$} is
below the global value  \mbox{$\Omega_m=0.28$}, and in large
volumes it tends to the asymptotic value of \mbox{$\Omega_{m,{\rm
loc}}=0.08\pm0.02$}. The error of this value is mainly determined
by the measurement errors  of the optical radial velocities of
galaxies.

\section{THE PROBLEM OF MISSING DARK MATTER}

The observational fact that the virial masses of groups and
clusters of galaxies are not able to provide the global density
  \mbox{$\Omega_m=0.28$} was known in the literature for
quite a while.  This way, Tully~\cite{Tul1987:Karachentsev} has
used the data on 2367 galaxies with radial velocities below 3000
km/s for grouping the galaxies, applying the ``hierarchical tree''
technique. This   technique took into account the luminosity
difference of galaxies and united into groups about two thirds of
all the galaxies considered. The total contribution of virial
masses to $\Omega_m$ according to
Tully~\cite{Tul1987:Karachentsev} is exactly \mbox{$\Omega_{m,{\rm
loc}}=0.08$}. Vennik \cite{Ven1984:Karachentsev} and
Magtesian~\cite{Mag1988:Karachentsev} applied several other
algorithms for clustering the galaxies in the Local universe, and
obtained the following estimates: \mbox{$\Omega_{m,{\rm
loc}}\simeq 0.08$} and \mbox{$\Omega_{m,{\rm loc}}\simeq 0.05$},
respectively.  Both authors took into account the individual
luminosities of  galaxies.

 Crook et al.
\cite{Crook2007:Karachentsev} used the percolation method  (FoF)
to identify the groups of galaxies with apparent magnitudes
\mbox{$K<11.25$}  from the 2MASS catalog. Fifty-three percent of
all galaxies were clustered  at the average density contrast of
\mbox{$\Delta\rho/\rho\sim80$}.  Their total contribution in
$\Omega_m$ amounted to \mbox{$0.10$--$0.13$} depending on the
method of virial mass estimation. The differences in luminosity
of galaxies   were ignored here.  Trying to ``reach out'' to the
global value of  \mbox{$\Omega_m=0.28$}, Crook et
al.~\cite{Crook2007:Karachentsev} made another version of the
group catalog with a softer condition for the average density
contrast $\Delta\rho/\rho\sim12$. This technique fitted 73\% of
galaxies into groups and clusters, and the contribution of galaxy
systems in $\Omega_m$ amounted to \mbox{$0.14$--$0.23$}. However,
most of these low-contrast aggregates are not virialized systems,
since their crossing time is comparable with the age of the
Universe or even exceeds it.

Bahcall et al. \cite{Bah2000:Karachentsev} considered the
contribution  to  $\Omega_m$ of systems of galaxies at all
scales---from pairs to superclusters.  Assuming that the main
contribution to  $\Omega_m$ is introduced by rich clusters, the
authors obtained the value of $\Omega_m=0.16\pm0.05$. However,
despite an increase in the average virial mass-to-luminosity
ratio from pairs and groups to clusters and superclusters, the
main contribution to  $\Omega_m$ is still made by the small
systems like the Local Group. Only 2\% of the $K$-band luminosity
falls on the  clusters richer than Virgo
\cite{Eke2005:Karachentsev}, and the relative fraction of virial
mass in them does not exceed 10--15\%. Note that in the volume we
consider ($V_{\rm LG}<3500$ km/s) the relative contribution of
the Virgo cluster in the total mass of the Local universe is
about 15\%, i.e. it is not determinative on the background of
less populated systems.

Thus, the most refined methods of estimating the virial mass in
systems of different size and population lead to the value of the
local  \mbox{$(D\leq 50$~Mpc)} average  density of matter of
\mbox{$\Omega_{m,{\rm loc}}=0.08\pm0.02$},  what is \mbox{3--4}
times lower than the global value of \mbox{$\Omega_{m,{\rm
glob}}=0.28\pm0.03$}  in the standard $\Lambda$CDM
cosmology~\cite{Fuk2004:Karachentsev,Spre2007:Karachentsev}.
Various possible explanations of this contradiction were proposed
in the literature. We shall list three of them here.

1) Dark matter in the systems of galaxies extends far beyond
their virial radius, so that the total mass of a group/cluster is
3--4 times larger than the virial estimate.

2) The diameter of the considered region of the Local universe,
90~Mpc, does not correspond to the true scale of the ``homogeneity
cell''; our Galaxy may be located inside a giant void sized about
100--500~Mpc, where the mean density of matter is 3 to 4 times
lower than the global value.

3) Most of the dark matter in the Universe, or about two thirds
of it is not associated with groups and clusters of galaxies, but
distributed in the space between them in the form of massive
dark clumps or as a smooth ``ocean.''

Let us discuss each of these hypotheses in more detail.

\subsection{Dark Halos around Groups and Clusters}

Tavio et al. \cite{Tav2008:Karachentsev} and Masaki et al.
\cite{Mas2011:Karachentsev} considered the integral distribution
of  dark halo mass as a function of distance, expressed in the
units of virial radius. Assuming a standard NFW-profile of the
halo,  they estimated that about 50\% of the total mass
is located  outside the virial radius. According to Rines
and  Diaferio \cite{Rin2006:Karachentsev},  the collapse region
surrounding a galaxy cluster between the radii of
$R_{\rm VIR}$ and $R_0\simeq3.7R_{\rm VIR}$  may be containing a
mass of \mbox{$(1.19\pm0.18)M_{\rm VIR}$}.

On the other hand, the modelling of orbits of test particles
inside a massive halo, performed in~\cite{Ander2011:Karachentsev}
made these authors conclude that the ratio of the total halo mass
to the virial mass is on the average only $M_T/M_{\rm VIR}=1.25$.
We consider this result to be more realistic. As we noted in the
introduction, the total mass estimates of the Local Group and
other nearby groups within a radius of $R_0$ are in good
agreement with their virial masses. A similar correspondence
(with an error of 30--50\%) was also found  in the closest to us
Virgo and Fornax
clusters~\cite{Kar2010:Karachentsev,Nas2011:Karachentsev}. In any
case, the assumption of the presence   between   $R_{\rm VIR}$ and
$R_0$ of a mass 2 to 3 times larger than   $M_{\rm VIR}$ clearly
contradicts  the existing observational data.

\subsection{Extended Local Void}

The idea that our Galaxy is located near the center of a vast
cosmic void \cite{Shaf2009:Karachentsev,Rom2010:Karachentsev} was
repeatedly set forth as an alternative to the model of accelerated
expansion of the Universe, created based on observations of
distant Supernovae. However, as indicated by the data in Figs.~3
and 4, the location of our Galaxy is characterized by an excess,
rather than by a deficiency of  local density at all scales up to
45~Mpc. It can be assumed that the local density excess elevates
above the extended surrounding void  like the central peak of a
lunar crater. Apart from its slight artificiality, this
assumption is in contradiction with observational data. The
counts of galaxies in the $K$-band, made up to the deep limit in
different directions
\cite{Djor1995:Karachentsev,Ber1998:Karachentsev,Tot2001:Karachentsev,Huang2001:Karachentsev}
do not show any signs of the existence of an extensive local void
sized about \mbox{100--500~Mpc.}

\subsection{The Population of Dark Attractors}

For obvious reasons, the hypothesis of the existence between galaxy
groups and clusters of a large number of invisible dark halos with
different masses is difficult to prove observationally.
Nevertheless, there already exist some evidence in favor of such
an unusual picture. Karachentsev et
al.~\cite{Kar2006:Karachentsev} drew attention to the presence
among the isolated galaxies listed in the KIG catalog
\cite{Kara1973:Karachentsev} of a fraction of objects with heavily
distorted structures. A good example is the \mbox{KIG~293 $=$
UGC~4722}  galaxy with an extended curved tail (Fig.~5). It is
commonly assumed  that the structures like these are formed during
a tight interaction of galaxies of an approximately equal mass.
However, in the broad vicinity of UGC~4722 there are no neighbors
that would be able to produce the observed tidal perturbations. As
a probable explanation for this phenomenon we may assume here a
case of interaction of a normal galaxy with an invisible dark
object having a mass of  about $10^9~M_{\odot}$. Recently,
a similar population of distorted isolated  galaxies was found in
the  Local Orphan Galaxy catalog \cite{Kar2011b:Karachentsev}. In
both samples, KIG and LOG, the relative number of such objects is
small, \mbox{about 4\%.}

\begin{figure}[tbp!!!]
\setcaptionmargin{5mm} \onelinecaptionstrue
\includegraphics[width=\columnwidth]{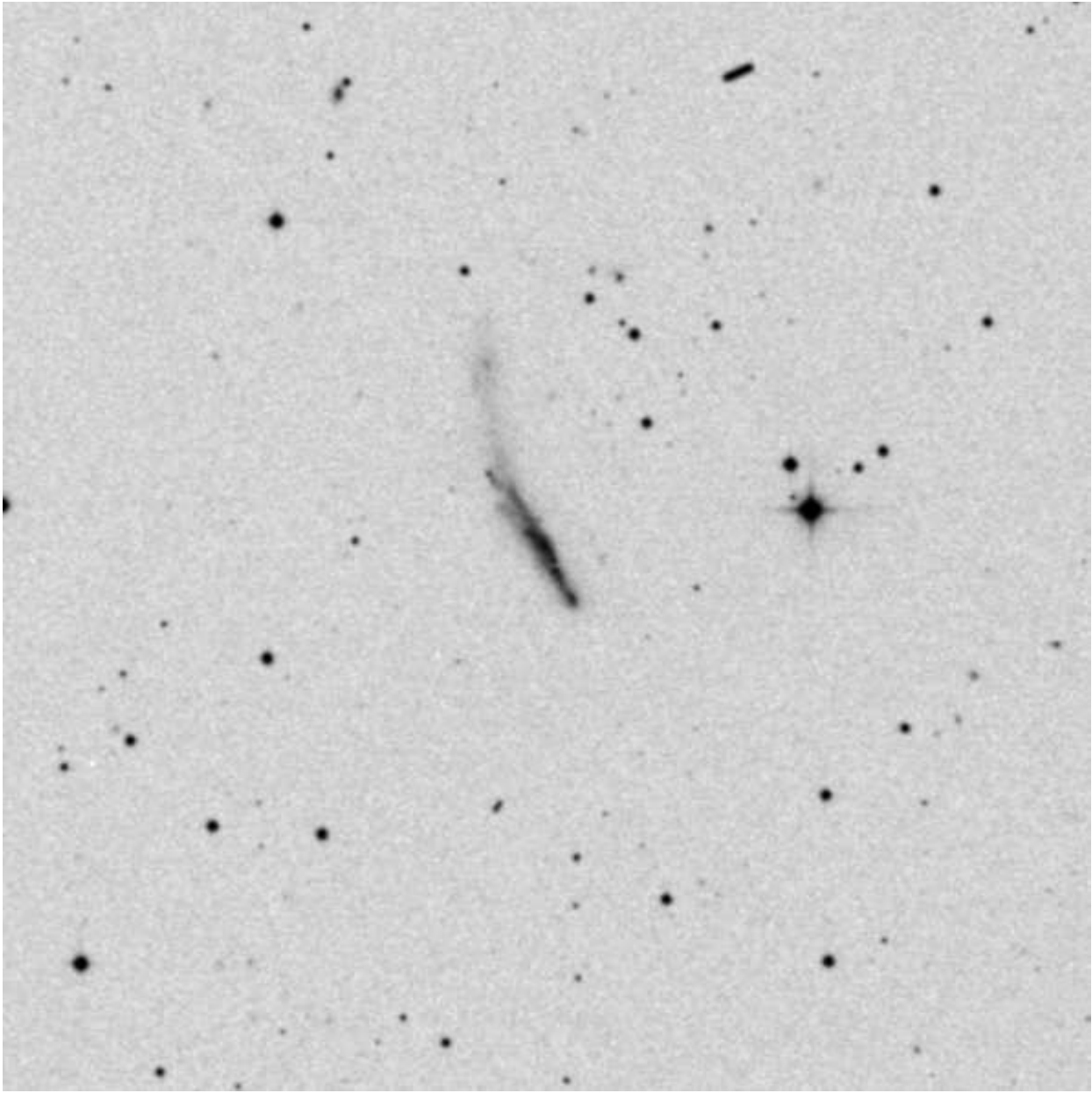}
\captionstyle{normal} \caption{An isolated galaxy UGC~4722 with
signs of a strongly perturbed structure. The image size is
\mbox{$10'\times10'$}. The short bold strip in the upper right is a
trace of an asteroid.}
\end{figure}

Examining the structures of the Einstein gravitational lenses with
the Hubble Space Telescope, Vegetti et
al.~\cite{Veg2010:Karachentsev,Veg2012:Karachentsev} have
discovered at the cosmological distances of   $z=0.22$ and
\mbox{$z=0.88$} two cases, indicating the presence of invisible
companions around the lens galaxies. According to the authors, the
masses of these dark companions amount to
$10^8$--$10^9~M_{\odot}$, and the lower limit of their
mass-to-luminosity ratio exceeds \mbox{$120~M_{\odot}/L_{\odot}$}.
It is as yet viewed difficult to evaluate the cosmic abundance of
such dark substructures.

Shan et al.~\cite{Shan2011:Karachentsev} investigated the effects
of weak gravitational lensing of distant galaxies in an area sized
72$\Box\degr$, recorded at the CFHT telescope with subarcsecond
images. On the map of gravitational potential peaks,
reconstructed from these data, the authors have found 301 peaks,
126 of which were identified with the optical or X-ray clusters of
galaxies. About 60\% of the peaks remained unidentified, which may
indicate the existence at the epoch of  \mbox{$z\sim0.4$} of the
population of dark attractors with masses, typical of rich
clusters of galaxies. Similar observational arguments in favor of
the existence of massive dark clumps were given earlier by
Natarajan and Springel~\cite{Nat2004:Karachentsev} and Jee et
al.~\cite{Jee2005:Karachentsev} based on the effects of weak
lensing.

\begin{figure*}[tbp!!!]
\setcaptionmargin{5mm} \onelinecaptionstrue
\includegraphics[scale=0.8]{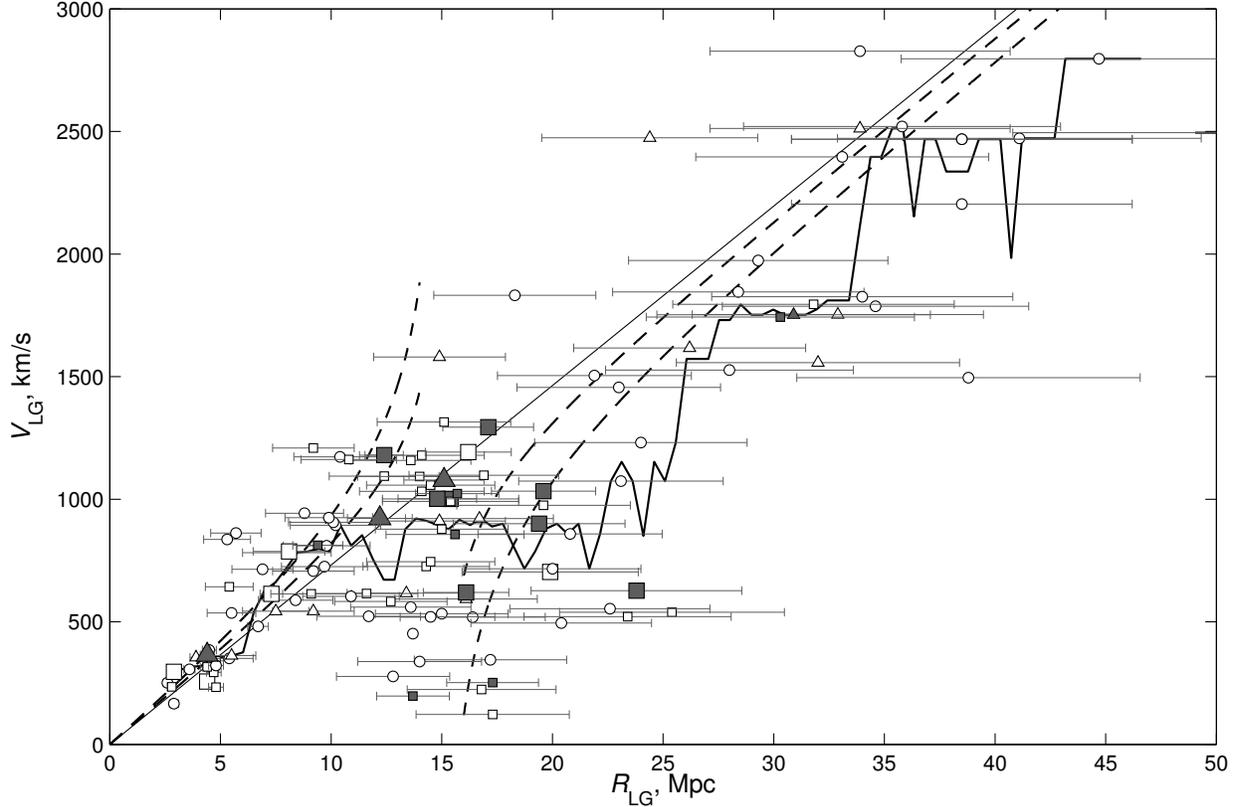}
\captionstyle{normal} \caption{The velocity--distance relation  for
122 galaxies in the Coma\,I region. The straight line corresponds
to the unperturbed Hubble flow with the parameter
$H_0=73$~km/s/Mpc. The squares mark the group members, the
triangles denote the members of triplets and pairs, and the
circles---single galaxies. The galaxies with bulges (E, S0, Sa)
are shown by dark symbols. The horizontal segments correspond to
the distance measurement errors. The broken line describes the
behavior of the running median with a window of 2~Mpc. Two dashed
lines correspond to the pattern of galaxy infall on a point-like
attractor with a mass of  \mbox{$0.5\times10^{14}~M_{\odot}$} and
$2.0\times10^{14}~M_{\odot}$, located at a distance of 15~Mpc, for
the case where the line of sight passes through the center of the
attractor.}
\end{figure*}

Among the galaxies of the Local universe with  measured distances
and radial velocities there exist galaxies with large negative
peculiar velocities. Inspecting these cases, Karachentsev et
al.~\cite{Kar2011a:Karachentsev} noted the fact that a half of
these rapidly moving galaxies are concentrated in a small region
of the sky  \mbox{(${\rm RA}\simeq12^h20^m, {\rm
Dec}\simeq+30^{\circ}$),} called the Coma\,I \linebreak cloud.
This region contains some scattered groups with the distances of 4
to 30~Mpc. The Hubble diagram for the Coma\,I galaxies (Fig.~6)
reveals systematic deviations from the linear relation $V=H_0R$,
characteristic of the infall of galaxies into the attractor. The
effect of infall is consistent with the observational data at the
attractor mass amounting to around $2\times10^{14}~M_{\odot}$ at
a distance of about 15~Mpc from us. The total luminosity of
galaxies in the zone of the assumed attractor is
\mbox{$L_K=1.0\times10^{12}~L_{\odot}$,} which gives the
mass-to-luminosity ratio of
\mbox{$M/L_K\sim200~M_{\odot}/L_{\odot}$}. This $M/L_K$ ratio is 4
times higher than that found in rich clusters like Coma. It is
possible that we have come across the first case of a nearby dark
attractor having a mass, typical of clusters of galaxies.

%

It should be emphasized that the mean local matter density
 $\Omega_m$ can be determined not only by summing the
virial masses of galaxy systems, but also from the analysis of the
peculiar velocity field at sufficiently large scales. Abate and
Erdogdu~\cite{Abate2009:Karachentsev} investigated the peculiar
velocity field of galaxies from the SFI++
sample~\cite{Sprin2009:Karachentsev} and  at the scale of about
\mbox{$6000$~km/s} obtained the estimate of
\mbox{$\Omega_m=0.09$--$0.23$}. This value should obviously
include the dark matter between the groups and clusters, if it
exists.

A similar analysis of the SFI++ sample was made by Davis et
al.~\cite{Dav2011:Karachentsev} involving the data on redshifts
of galaxies from the 2MASS survey~\cite{Jar2000:Karachentsev}. The
authors deduced that the local field of peculiar velocities is in
a good agreement with the local topography of gravitational
potential (the distribution of dark matter follows the
distribution of galaxies). However, Davis et al. specify that the
smooth component of dark matter can not be tested by the means of
this technique. Another local estimate
\mbox{$\Omega_m=0.20\pm0.07$} at the scale of approximately
\mbox{$100$~Mpc} is obtained in~\cite{Bil2011:Karachentsev} based
on the behavior of the clustering dipole of  2MASS galaxies
relative to the direction of motion of the Local Group in the
cosmic microwave background radiation frame.

\section{DYNAMIC COMPONENTS OF THE LARGE-SCALE STRUCTURE}

\begin{figure*}[tbp!!!]
\setcaptionmargin{5mm} \onelinecaptionstrue
\includegraphics[scale=0.9,angle=-90]{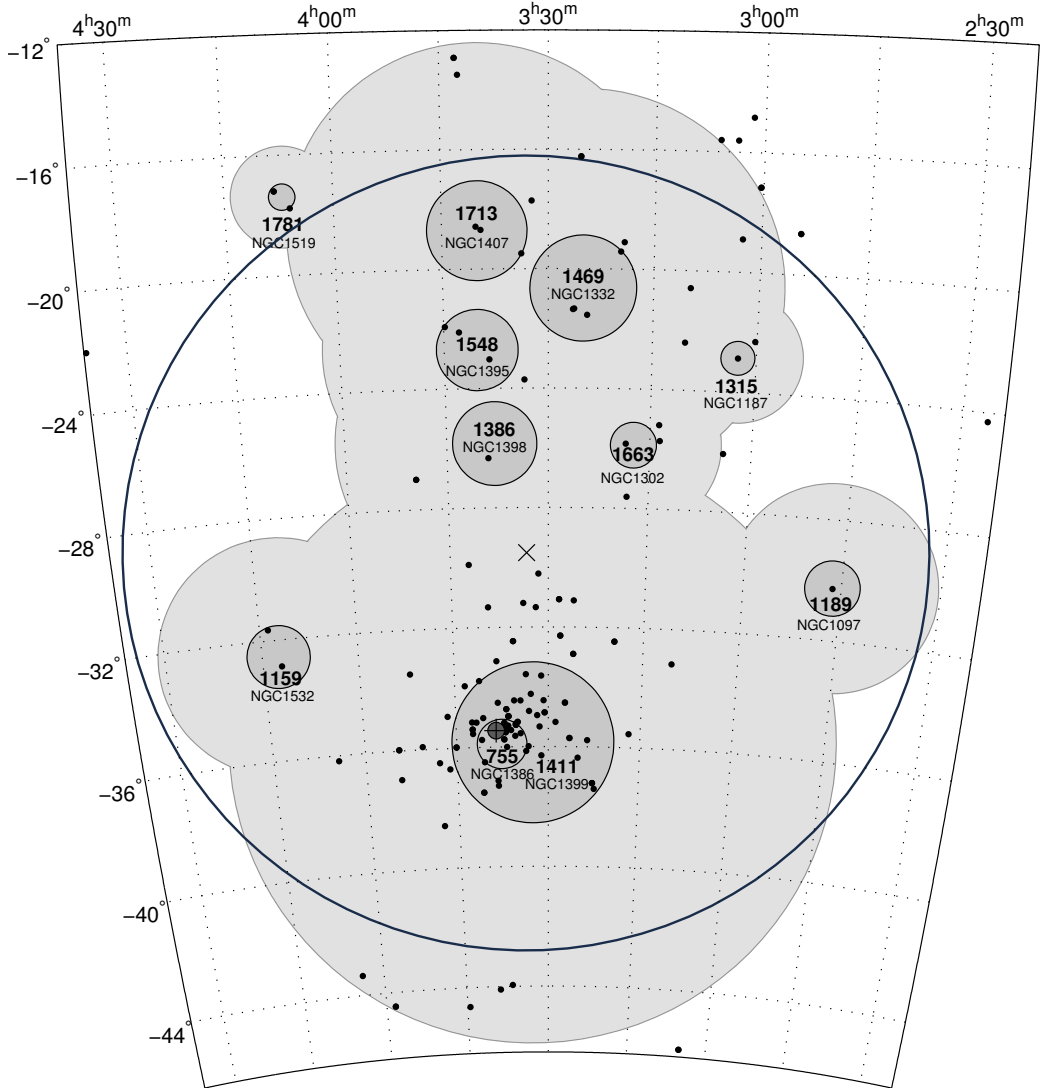}
\captionstyle{normal} \caption{ The Fornax cluster and Eridanus groups of
galaxies in equatorial
coordinates. The dark circles correspond to the virial radii of
the systems, the lighter circles correspond to the
zones of collapse around them. The NGC numbers mark the brightest
members of each group, and the figures in bold---the mean radial
velocities of groups in km/s.}
\end{figure*}

The following elements of the large-scale structure of the
Universe are typically singled out based on their geometric
features: large and small cosmic voids, filaments and walls
bordering the voids, clusters of galaxies in the nodes at
intersections of the filaments and walls. As the Universe expands, the
density perturbations increase and the elements of the large-scale
structure become more contrasting, which is clearly illustrated by
numerous N-body simulations.

From the dynamic point of view, it is reasonable to divide the
elements of the large-scale structure into three categories:

a) virialized zones of groups and clusters, where the balance
between the kinetic and potential energy  ($2T+U=0$) was
established, and the galaxies have ``forgotten'' the initial
conditions of their formation;

b) collapsing regions around the  virialized zones, limited by
the zero velocity spheres with the $R_0$ radius;

c) the remaining, infinitely expanding regions of the ``general
field'' which comprise the population of voids and diffuse
filaments.

An example of such a separation is shown in Fig.~7, which
represents a complex of nearby groups in the Fornax-Eridanus
constellations in equatorial
coordinates~\cite{Nas2011:Karachentsev}. Here the virialized
regions of the Fornax cluster (located downward from the center of
the figure) and the groups, adjacent to each other in the Eridanus
are marked by dark circles. The lighter circles correspond to the
$R_0$ radii around the groups. It is clear from the figure that
the spheres of $R_0$ radius mutually intersect, forming a surface,
under which the Eridanus groups and the Fornax cluster will merge
into a single dynamic aggregate over time.


\begin{table}
\setcaptionmargin{0mm} \onelinecaptionstrue
\caption{Basic parameters of three dynamic regions in the large-scale structure}
\begin{tabular}{l|c|c|c} \hline
\multicolumn{1}{c|} {Basic parameters} & Virialized  &Collapsing &Remaining \\
 & zones           &regions        &expanding background\\
 \hline
Relative number of galaxies & 54\%   & $\sim20$\% & $\sim26$\% \\
Relative amount of stellar mass  & 82\%   & $\sim8$\%  &   $\sim10$\% \\

Relative fraction of occupied volume   & 0.1\%  & 5\%     &         95\% \\
Contribution to $\Omega_m$  & 0.06   & 0.02       &         0.20 \\
Mean ratio of dark to stellar mass, $M_{\rm DM}/M_*$  & $\sim26$ &$\sim87$&    $\sim690$\\
\hline
\end{tabular}
\end{table}

Based on the results of clustering of galaxies in the Local
universe~\cite{Mak2011:Karachentsev} and other observational data
on similar systems, it is possible to identify a number of
important parameters, characterizing the dynamic status of three
components of the large-scale structure at the current epoch.
These parameters are listed in the table. Some of them as yet
contain a significant uncertainty.

As follows from the first two rows in the table, more than a half
of all galaxies and more than four fifths of their stellar mass
are already located within the virial zones. The subsequent
infall of galaxies from the surrounding collapsing regions will
bring a relatively small addition to the existing virial masses.
In this sense we can say that the main stage of dynamical
evolution of the large-scale structure is already completed.

The third row of the table shows that the virialized volumes of
groups and clusters occupy only 0.1\% of the total volume, while
the collapse regions around them take up about 5\% of the volume
of Universe. The remaining 95\% of the volume, related to the
general field contain only about 10\% of stellar mass. Thus, the
contrast of the mean stellar mass density  between the virialized
volumes and the general field reaches about 7000.

The penultimate row of the table demonstrates the contribution of
three different dynamic zones to $\Omega_m$, considering that
overall they yield a default value of  $\Omega_{m,{\rm
glob}}=0.28$. It is interesting to note here that the contrast of
the average densities of dark matter between the virialized zones
and the general field is \mbox{about 280.}

Finally, the last line presents the characteristic ratio of the
dark matter mass to the luminous (stellar) mass in the three
designated regions.  As above, the  stellar
mass of the galaxy $M_*$ is expressed in terms of its $K$-band
luminosity, assuming  that \mbox{$M_*/L_K\simeq
1.0~M_{\odot}/L_{\odot}$ \cite{Bell2003:Karachentsev}}.

One important circumstance should be noted here. The algorithm,
applied in \cite{Mak2011:Karachentsev}  for the selection of
groups assumes that in all galaxies, irrespective of their
luminosity, type or surrounding density, the ratio of the total
mass of the halo to the mass of the stellar component has one and
the same value, \mbox{$M_T/M_* =6$}. This dimensionless quantity
is the only more or less arbitrary parameter of the applied
algorithm (as opposed to the percolation algorithm FoF, where the
choice of two arbitrary parameters is required: the maximal
difference in radial velocities and the maximal projected
separation of the components of a virtual pair). According to the
``Bolshoi'' N-body simulations within the $\Lambda$CDM model
\cite{Tru2011:Karachentsev}, the \mbox{$M_T/M_*=6$} ratio should
approximately be  met (within the \mbox{$\pm1\sigma$} band) for
all the galaxies with stellar masses in the range of
$\log(M_*/M_{\odot})=[8.5$--$11.0]$, which indirectly justifies
the choice of the  single clustering parameter $M_T/M_*$.

Thus, using the initial value of  $M_T/M_*=6$ for individual
halos of galaxies, for the virialized regions of groups and
clusters we get the mean ratio of  $M_T/M_*\simeq26$ (it
systematically grows in the transition from pairs and groups to
clusters), and the remaining space of the general field is
characterized by the ratio $M_T/M_*\sim690$, i.e. two orders
higher than that of a typical galaxy.

\section{CONCLUDING REMARKS}

As it is widely known, some predictions of the standard
$\Lambda$CDM cosmological model are not well consistent with the
observational data available nowadays. First and foremost, this
concerns the problem of ``missing satellites"
~\cite{Klyp1999:Karachentsev}, which lies in the fact that the
theory predicts the number of companions in the Milky Way-type
galaxies to be dozens of times greater than that observed around
our Galaxy, M\,31, and other neighboring high-luminosity
galaxies. Various explanations of this inconsistency have been
proposed, based on some features of the star formation process in
dwarf galaxies at the  $z\sim10$ epoch. However, the problem of
``missing satellites''  still remains.

Another puzzle is that the theory of formation of chemical
elements in the hot expanding Universe gives the value of cosmic
baryon abundance of  \mbox{$\Omega_b=0.045\pm0.005$
\cite{Fuk2004:Karachentsev}.} However, the current observational
data reveal only 1/10 of these baryons, existing in the form of
stars and gas in the galaxies. It is assumed that the bulk of the
baryons may be distributed between the galaxies alike the warm
\mbox{($T\sim10^5$~K)}, non-virialized
``broth''~\cite{Cen1999:Karachentsev}. There have been reports in
the literature on the likely observational detection of ``missing
baryons'' \cite{Stok2006:Karachentsev,Nar2010:Karachentsev}.
Nevertheless, this problem can neither be considered definitively
resolved.

A significant divergence between the local \linebreak
\mbox{($0.08\pm0.02$)} and global  \mbox{($0.28\pm0.03$)} values
of the average matter density  adds yet another mystery, the
problem of ``missing dark matter''. Unlike the situation with
missing companions and baryons, the lack of dark matter in the
Local universe is not characterized by one or two orders of
magnitude, but only by a factor of 3, which, however, is quite a
few for the so-called ``precision cosmology era''. Contradictory
estimates of $\Omega_m$ probably indicate that the assumption of
proportional distribution of dark and stellar matter,
$\log(\rho_{\rm DM})\propto\log(\rho_*)$, though being
convenient, is however not quite a justifiable paradigm. In other
words, our Universe might happen to be more hidden and dark than
we thought until recently.

\begin{acknowledgments}
I thank James Peebles for valuable comments.
This work was supported by the grant of the \linebreak Russian
Foundation for Basic Research (grant \linebreak
no.~\mbox{11-02-90449-Ukr-f-a}) and by the state contract
``Cosmology of Nearby Universe'' no.~14.740.11.0901.
\end{acknowledgments}

\end{document}